\documentclass[journal=cmatex,manuscript=article]{achemso}

\usepackage{amsmath,amssymb}
\usepackage{gensymb}
\usepackage{textcomp}
\usepackage{graphicx}% Include figure files
\DeclareGraphicsExtensions{.eps,.bmp}
\usepackage{chemformula} % Formula subscripts using \ch{}
\usepackage[T1]{fontenc} % Use modern font encodings
\usepackage{bm}% bold math
\usepackage{hyperref}% add hypertext capabilities

\newcommand{\BaFeAl}{BaFe$_2$Al$_9$}

\author{William R. Meier}
\email{4wm@ornl.gov}
\affiliation{Materials Science \& Technology Division, Oak Ridge National Laboratory, Oak Ridge, Tennessee 37831}%

\author{Bryan C. Chakoumakos}
\affiliation{Neutron Scattering Division, Oak Ridge National Laboratory, Oak Ridge, Tennessee 37831}%

\author{Satoshi Okamoto}
\affiliation{Materials Science \& Technology Division, Oak Ridge National Laboratory, Oak Ridge, Tennessee 37831}%

\author{Michael A. McGuire}
\affiliation{Materials Science \& Technology Division, Oak Ridge National Laboratory, Oak Ridge, Tennessee 37831}%

\author{Rapha\"el P. Hermann}
\affiliation{Materials Science \& Technology Division, Oak Ridge National Laboratory, Oak Ridge, Tennessee 37831}%

\author{German D. Samolyuk}
\affiliation{Materials Science \& Technology Division, Oak Ridge National Laboratory, Oak Ridge, Tennessee 37831}%

\author{Shang Gao}
\affiliation{Materials Science \& Technology Division, Oak Ridge National Laboratory, Oak Ridge, Tennessee 37831}
\alsoaffiliation{Neutron Scattering Division, Oak Ridge National Laboratory, Oak Ridge, Tennessee 37831}%

\author{Qiang Zhang}
\affiliation{Neutron Scattering Division, Oak Ridge National Laboratory, Oak Ridge, Tennessee 37831}%

\author{Matthew B. Stone}
\affiliation{Neutron Scattering Division, Oak Ridge National Laboratory, Oak Ridge, Tennessee 37831}%

\author{Andrew D. Christianson}
\affiliation{Materials Science \& Technology Division, Oak Ridge National Laboratory, Oak Ridge, Tennessee 37831}%

\author{Brian C. Sales}
\email{salesbc@ornl.gov}
\affiliation{Materials Science \& Technology Division, Oak Ridge National Laboratory, Oak Ridge, Tennessee 37831}%

\title[CDW in BaFe$_2$Al$_9$]{A catastrophic charge density wave in BaFe$_2$Al$_9$}

\abbreviations{CDW,DFT}

\begin{document}

\begin{tocentry}
	
\includegraphics[width=3.25in]{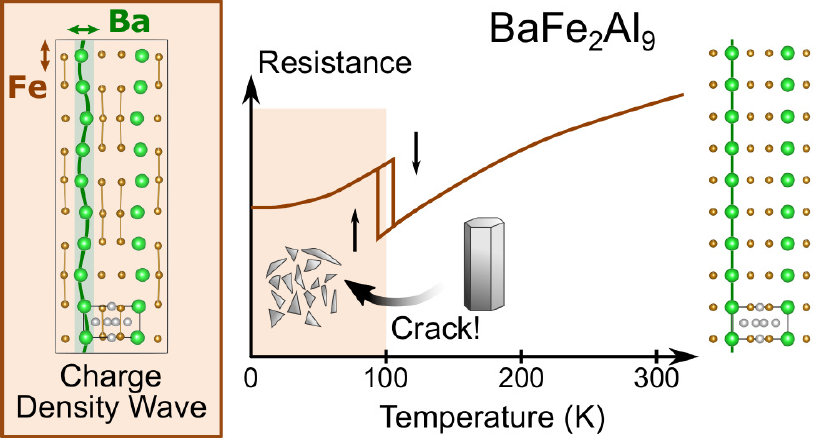}

\end{tocentry}
\begin{abstract}
Charge density waves (CDW) are modulations of the electron density and the atomic lattice that develop in some crystalline materials at low temperature. We report an unusual example of a CDW in BaFe$_2$Al$_9$ below 100\,K. In contrast to the canonical CDW phase transition, temperature dependent physical properties of single crystals reveal a first-order phase transition. This is accompanied by a discontinuous change in the size of the crystal lattice. In fact, this large strain has catastrophic consequences for the crystals causing them to physically shatter. Single crystal x-ray diffraction reveals super-lattice peaks in the low-temperature phase signaling the development of a CDW lattice modulation. No similar low-temperature transitions are observed in BaCo$_2$Al$_9$. Electronic structure calculations provide one hint to the different behavior of these two compounds; the d-orbital states in the Fe compound are not completely filled. Iron compounds are renowned for their magnetism and partly filled d-states play a key role. It is therefore surprising that BaFe$_2$Al$_9$ develops a structural modulation at low temperature instead of magnetic order.
%BaFe$_2$Al$_9$ offers many opportunities for chemical substitution to tune this abrupt CDW transition and uncover why this lattice modulation is accompanied by such a destructively large lattice change. 

\end{abstract}

This manuscript has been authored by UT-Battelle, LLC under Contract No. DE-AC05-00OR22725 with the U.S. Department of Energy. The United States Government retains and the publisher, by accepting the article for publication, acknowledges that the United States Government retains a non-exclusive, paid-up, irrevocable, world-wide license to publish or reproduce the published form of this manuscript, or allow others to do so, for United States Government purposes. The Department of Energy will provide public access to these results of federally sponsored research in accordance with the DOE Public Access Plan (http://energy.gov/downloads/doe-public-access-plan).
\section{Introduction}
\label{sec:Intro}

\begin{figure}
\includegraphics[width=3.33in]{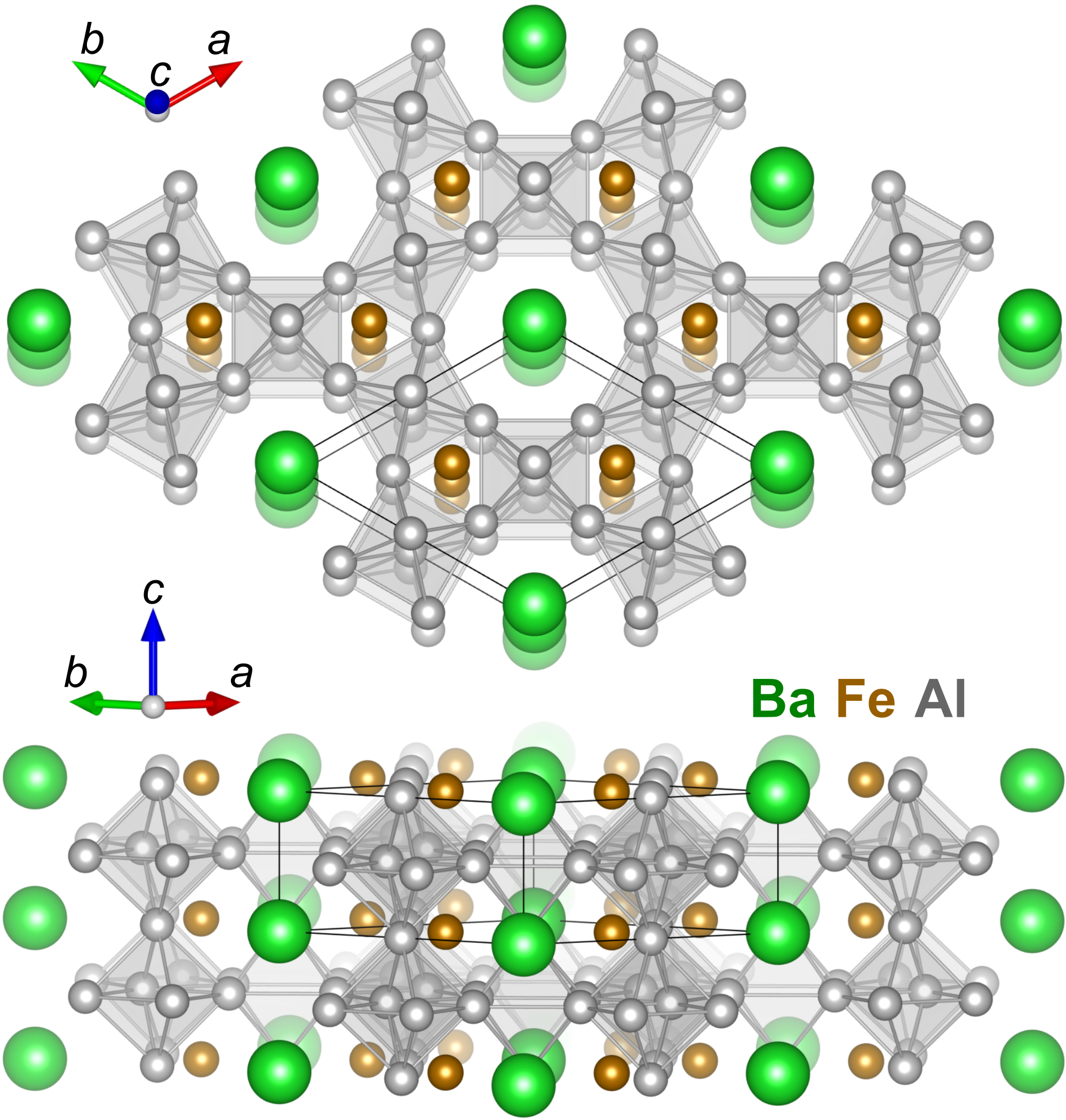}
\caption{\label{fig:Structure} 
	Crystal structure of BaFe$_2$Al$_9$ highlighting network of corner-sharing aluminum octahedra. Figures were generated in VESTA \cite{Momma2011_Vesta3}.
}
\end{figure}

Some crystalline materials experience rearrangements of their atomic structure when they are heated or cooled. These structural transitions are important in many technologies including hardening steel\cite{Bhadeshia2006_Steels}, battery materials\cite{Orikasa2013_LiFePO4-FePO4-PhaseChangeBatteries,Yahia2013_LiCoO2-PhaseTransition} and next generation electronic devices\cite{Tomforde2011_Ge-Sb-Te-PhaseChangeThinFilms,Bischoff2017_PhaseEngineeringNbSe2,Meddar2012_IncrePhaseTransTempMultiferroicMnWO4}. In addition, examining the details of the structural and property changes offers insights into the mechanisms at play and inspires design of new useful materials. A charge density wave is a curious variant of a structural transition.

A charge density wave (CDW) is a modulation of a material's atomic lattice that develops on cooling.
This phenomenon is characterized not only by a periodic redistribution of electronic density (the charge density wave), but also a complementary static periodic displacement of the atoms.
Critically, most CDWs have wavelengths distinct from the repeat unit of the original crystal lattice.\cite{Gruner1994_DensityWavesInSolids}
The simplest picture of a CDW is the Peierls model. Partially filled orbitals in a chain of atoms display an instability to forming alternating bonds.\cite{Burdett1995_ChemicalBondingInSolids,Gruner1994_DensityWavesInSolids,Hoffmann1987_ChemistryPhysicsMeetInTheSolidState} This is a powerful yet simple idea often used to understand dimers and distorted networks in a wide variety of chemical systems.\cite{OwensBaird2019_NiP2-polymorphs,Kobayashi2019_TrimerFormation1T-CrSe2,Alemany2009_HostGuestInteractionsCa7N4Mx}
Although CDWs are usually introduced conceptually with the Peierls model, the origin of CDW order is not always clear.\cite{Gruner1994_DensityWavesInSolids,Zhu2017_MisconceptionsChargeDensityWaves,Gruener1988_DynamicsOfCDWs,Johannes2008_NestingAndChargeDensityWaves}

CDW materials host unusual physical properties including metal-insulator transitions and nonlinear electrical conductivity.\cite{Gruner1994_DensityWavesInSolids} In addition, a number of CDW systems can be modified to suppress the modulation and induce superconductivity.\cite{Chikina2020_CDW->SC_PdxTaSe2,Chen2016_CDWinStronglyCorrelatedSystems} CDW order has taken on new importance in the field of 2D materials like the transition metal dichalcogenides \cite{Chen2016_CDWinStronglyCorrelatedSystems,Duvjir2018_MetalInsulatorTrans+CDWinVSe2Monolayer,Malliakas2013_NbNbInteractionsAndCDWof2H-NbSe2,Duong2017_RamanOfCDWin1T-TiSe2,Wilson1975_CDW+SuperlatticesInTransitionMetalDichalcogenides,Rossnagel2011_OriginOfCDWsInTransitionMetalDichalcogenides,Bianco2019_CDWinNbS2at2DLimit}. CDW modulations can modify the electronic, magnetic and optical properties of these materials in useful ways.\cite{Fu2016_Thin-TaS2-TuneCDW,Feng2018_ElectronicStructure+CDWinMonolayerVSe2,Fumega2020_Controlled2DFerromagIn1T-CrTe2,Xi2015_EnhancedCDW-MonolayerNbSe2} Classic examples of CDWs occur in materials with 1D or 2D electronic structures derived from weakly-coupled chain-like (K$_{0.3}$MoO$_3$ and NbSe$_3$)\cite{Gruner1994_DensityWavesInSolids,Whangbo1984_NbSe3-CDW+BandStruct,Hor2005_NanowiresNanoribbonsCDW-NbSe3} or sheet-like motifs (NbSe$_2$ and EuAl$_4$)\cite{Wilson1975_CDW+SuperlatticesInTransitionMetalDichalcogenides,Weber2011_PhononCollapseAndOriginOfCDW-2H-NbSe2,Arguello2014_VisualizingCDW-2H-NbSe2,Shimomura2019_CDW-EuAl4}. Few examples of 3D-network compounds with CDWs are known.\cite{Zhu2017_MisconceptionsChargeDensityWaves}

\label{sec:Intro_129Family}
BaFe$_2$Al$_9$ is a member of a family of aluminum-rich, intermetallic compounds; $A$$M_2$Al$_9$. \citeauthor{Turban1975_BaFe2Al9-StructureType} introduced the family with Ba$M_2$Al$_9$ ($M$ = Fe, Co, Ni) and Sr$M_2$Al$_9$ ($M$ = Co, Ni).\cite{Turban1975_BaFe2Al9-StructureType} CaCo$_2$Al$_9$ \cite{Manyako1988_Ca-CoNi-Al-System700K} and EuCo$_2$Al$_9$ \cite{Thiede1999_EuCo2Al9} share the same structure and three indium-rich compounds are variations on the same theme; KCo$_2$In$_9$, KNi$_2$In$_9$ \cite{Lei2009_Eu3Co2In15+KCo2Al9+KNi2Al9} and BaIr$_2$In$_9$ \cite{Calta2015_BaIr2In9ThermalExpansion}.
\label{sec:Intro_129Structure}
The hexagonal structure of BaFe$_2$Al$_9$ ($P$6/$mmm$ No.~191) is presented in Fig.~\ref{fig:Structure}.\cite{Turban1975_BaFe2Al9-StructureType} First, note the 3D network of corner-sharing Al octahedra. \citeauthor{Vajenine1998_MagicElectronCountsAeM2Al9} discussed the network bonding in detail and noted that this aluminum framework provides some flexibility in the optimal electron count\cite{Vajenine1998_MagicElectronCountsAeM2Al9}. This likely explains why the structure forms with a series of transition-metals (Fe, Co, Ni) accommodating the additional electrons in the Al bands. Barium atoms are closely-spaced, forming columns along the $c$-direction within large channels in the Al-framework. The iron atoms are centered in a tri-capped triangular prisms of Al atoms. The nearest Fe-Fe (and Ba-Ba) spacings are unit cell height, $c$ around 3.93\,\AA\ at room temperature.
%At room temperature, Fe-Fe NN = $c$ around 3.93\,\AA. NNN = $a$/sqrt(3) about 4.63\,\AA
%Ba-Ba spacing in Ba metal 5.03\,\AA wikipedia. What about other Ba intermetallics? 
%BaAl4 distance Ba-Ba = 4.57\,\AA\ (http://dx.doi.org/10.1016/j.jallcom.2015.08.193) and 4.56\,\AA\ in BaGa4 (http://dx.doi.org/10.1016/j.jallcom.2015.08.193). Al sheets clearly dominate spacing here.

BaFe$_2$Al$_9$ came to our attention while searching the ICSD for transition-metal materials with kagome-lattice motifs.\cite{Bergerhoff1987_ICSD} Although Al (not Fe) formed the desired lattice, the compounds still held an opportunity for interesting magnetism. Specifically, the Materials Project database predicted that the electronic structure had partially filled d-orbitals\cite{Jain2013_MaterialProjectDatabase} from the stacked honeycombs of Fe. To our surprise, BaFe$_2$Al$_9$ develops a CDW without any signs of the magnetism we anticipated. 

We report a charge density wave below 100\,K in crystals of BaFe$_2$Al$_9$ grown from an aluminum melt. Physical property measurements reveal that this transition is first-order with a sufficiently large change of the lattice parameters to shatter the crystals (0.5 and 1.5\%). Single crystal x-ray diffraction resolves super-lattice peaks at low temperature signaling CDW order. In contrast, BaCo$_2$Al$_9$ has no phase transitions below room temperature. Electronic structure calculations reveal that the transition-metal d-orbitals are partly filled in BaFe$_2$Al$_9$ and completely filled in BaCo$_2$Al$_9$. Based on this comparison, we propose that d-electrons in the Fe compound play a key role in CDW formation. Iron is renowned for its magnetism and partly filled d-orbitals are key to its behavior. This makes it more surprising that BaFe$_2$Al$_9$ hosts CDW order. This dramatic structural transition is the first CDW in an Fe-intermetallic compound and a rare example of a CDW in a 3D-network compound. BaFe$_2$Al$_9$ has many opportunities for chemical substitution to tune this transition to investigate its origin and its concomitant, large lattice contraction.

\section{Experimental}
\label{sec:Experimental}

\subsection{Crystal growth}
\label{sec:Exp_growth}
Crystals of BaFe$_2$Al$_9$ and BaCo$_2$Al$_9$ were grown from an aluminum-rich melt using a starting atomic composition of Ba:$M$:Al = 6:8:86 ($M$\,=\,Fe and Co) in 1.5-2.1\,g batches. Barium pieces (Alfa Aesar 99.9\%), iron granules (Alfa Aesar 99.98\%), cobalt powder (Alfa Aesar 99.8\%) and aluminum shot (Alfa Aesar 99.999\%) were loaded into a 2\,mL alumina Canfield Crucible Set \cite{Canfield2016_CanfieldCrucibleSet}. The raw materials and crucibles were sealed in a fused silica ampoule filled with argon. In a box furnace, the assembly was heated to 1150\,\textdegree C over 6\,h, held for 12\,h then slowly cooled to 1000\,\textdegree C over 200\,h to grow the crystals. The ampoule was removed from the hot furnace, inverted into a centrifuge and spun to remove the remaining Al-rich solution from the crystals. 

\subsection{Products}
\label{sec:Exp_products}
This procedure yielded around 0.5\,g of metallic hexagonal crystals with habits ranging from needles to blocky columns 0.3-7\,mm in diameter which often spanned the interior of the crucible (up to 13\,mm long). Crystals of BaCo$_2$Al$_9$ were generally smaller and thinner. These brittle crystals exhibit conchoidal fracture with no evidence of cleavage. Flux on the crystal surfaces was dissolved away in a 1\,M HCl solution over 0.5-4\,h.

\subsection{Characterization}
\label{sec:Exp_characterization}

%\textbf{Physical properties.} 
Magnetization measurements were performed at 2\,kOe on single crystals or crushed crystals in plastic drinking straws with a Quantum Design Magnetic Property Measurement System (MPMS) between 2 and 300\,K using a rate of 1\,K/min. 4-probe electrical resistance was measured on crystals contacted with Pt wire and Ag epoxy (EPO-TEK H20E) between 2 and 300\,K (at 1\,K/min) with the ac-transport option of a 9\,T Quantum Design Physical Property Measurement System (PPMS). Small, previously shattered crystal fragments were more likely to hold together during resistance measurements. Heat capacity was measured using the Quantum Design heat capacity option of the PPMS using Apiezon N-grease between 1.8 and 200\,K.

%\textbf{Neutron diffraction} 
Powder neutron diffraction was performed using the POWGEN time of flight diffractometer at the Spallation Neutron Source at Oak Ridge National Laboratory\cite{Huq2019_POWGEN}. 1.83\,g of ground BaFe$_2$Al$_9$ crystals were held in a vanadium can in the POWGEN Automatic Sample Changer. Measurements were preformed at temperatures from 7 to 300\,K using neutron wavelength bands centered at 1.5 and 2.66\,\AA. Rietveld refinement was preformed using Jana2020.\cite{Petricek2014_Jana2006}

%\textbf{Powder XRD.} 
The lattice parameters were determined by powder x-ray diffraction (PANalytical X'pert Pro, with Cu-K$\alpha$1 radiation). An Oxford PheniX closed-cycle helium cryostat was used for scans below room temperature. Lattice parameters were determined from full 2$\theta$ scans taken at each temperature.

%\textbf{Single Crystal XRD.} 
Single-crystal diffraction data were collected from a 100\,\textmu m BaFe$_2$Al$_9$ crystal fragment as a function of temperature using a Rigaku XtaLAB PRO diffractometer with graphite monochromated Mo K$\alpha$ radiation ($\lambda$\,=\,0.71073\,\AA, 50\,kV and 40\,mA) equipped with a Rigaku HyPix-6000HE detector and an Oxford N-HeliX cryocooler. Peak indexing and integration were done using the Rigaku Oxford Diffraction CrysAlisPro software\cite{CrysAlisPRO}. An empirical absorption correction was applied using the SCALE3 ABSPACK algorithm as implemented in CrysAlisPro. Omega scans with counting times up to 60\,s per point ensured that the weak superlattice reflections could be easily seen. Structure refinements were performed in Jana2020.\cite{Petricek2014_Jana2006}

%\textbf{M\"ossbauer} 
Temperature-dependent $^{57}$Fe M\"ossbauer spectra were acquired on ground BaFe$_2$Al$_9$ crystals placed in a Janis SH-850 closed cycle cryostat. A 15 mCi $^{57}$Co source and a Ritverc-2 Tl@NaI detector were used with a Wissel drive. This was calibrated using a room-temperature alpha-iron sample, which serves as a reference for the isomer shift. Data was acquired between 295\,K and 10\,K, first on cooling and then on heating to reveal the hysteretic nature of the transition.

%\textbf{DFT-methods}
DFT calculations were carried out using the projector augmented wave method \cite{Kresse1999_ProjectorAugmentedWaveMethod} with the generalized gradient approximation in the parametrization of Perdew, Burke, and Enzerhof (PBE) \cite{Perdew1996_GGA} for exchange-correlation as implemented in the Vienna {\it ab}-\textit{initio} simulation package (VASP) \cite{Kresse1996_AbInitioEnergyCalculations}. For Ba, a pseudo-potential, in which semi-core s-states are treated as valence states, is used (Ba$_{sv}$ in the VASP distribution). For Fe, Co and Al standard potentials are used (Fe, Co and Al, respectively). For both cases, we use the experimental lattice parameters obtained at room temperature with a $12 \times 12 \times 12$ $\bf k$-point grid and an $E$ cutoff of 500~eV. Local Hubbard +$U$ correction is not included because both compounds are itinerant intermetallic systems. While the spin-orbit coupling was included, its effect is negligibly small because no heavy elements had states near the Fermi-level. In both cases, paramagnetic ground states are stabilized, consistent with experimental observations. 

\section{Results}
\label{sec:Results}

\begin{figure}
	\includegraphics[width=3.33in]{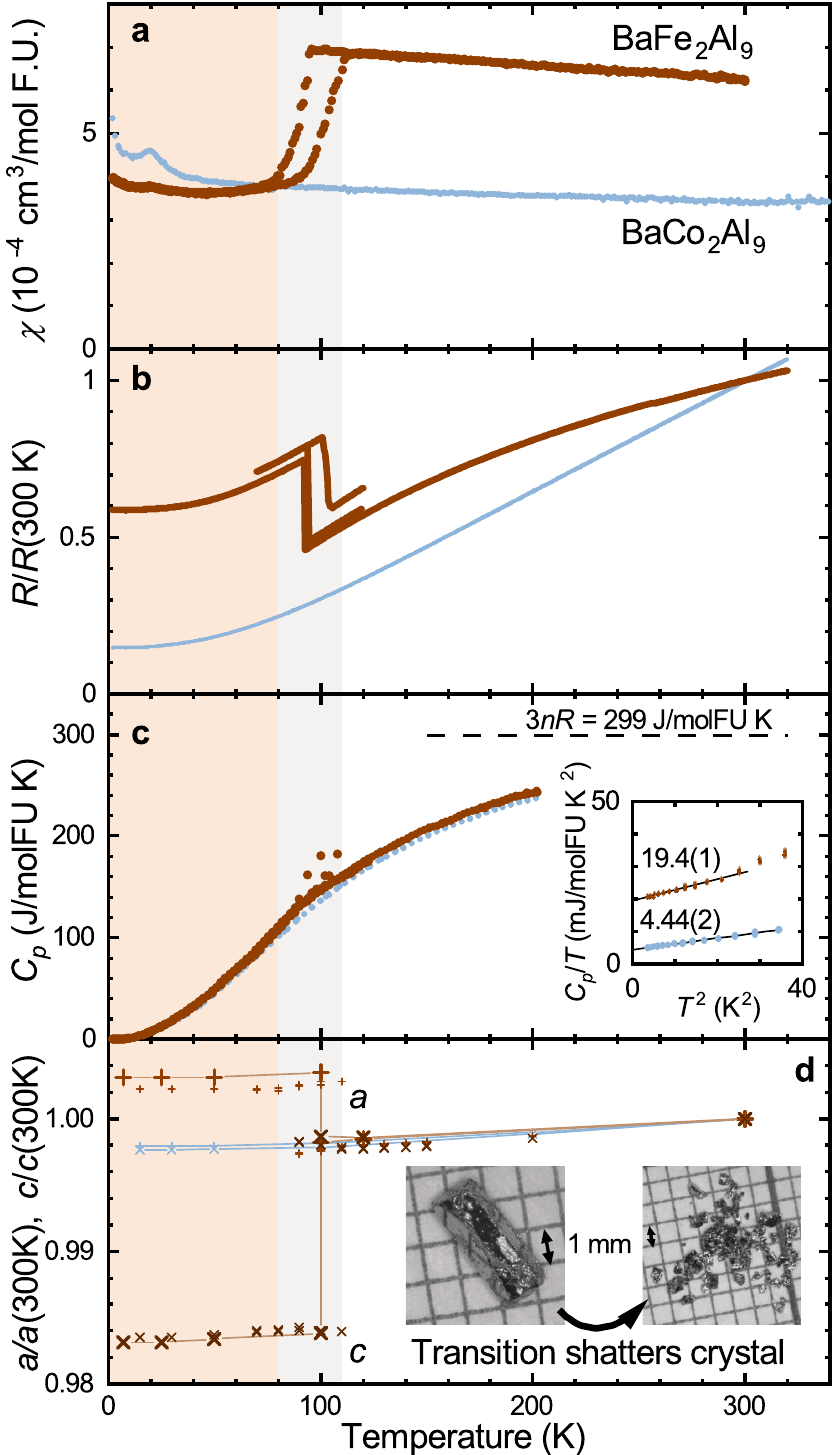}
	\caption{\label{fig:PhysicalProperties} 
		Physical properties of BaFe$_2$Al$_9$ (brown) and BaCo$_2$Al$_9$ (blue) reveal a first order phase transition in BaFe$_2$Al$_9$. (\textbf{a}) Magnetic susceptibility measured at 2\,kOe. (\textbf{b}) Electrical resistivity normalized to 300\,K value. (\textbf{c}) Specific heat capacity and inset showing estimated $\gamma$ in mJ/molF.U.\,K$^2$. (\textbf{d}) Relative a (+) and c ($\times$) lattice parameters of BaFe$_2$Al$_9$ determined by powder neutron diffraction (big symbols) and powder XRD (small symbols). The BaCo$_2$Al$_9$ data was measured by powder XRD. The inset shows a BaFe$_2$Al$_9$ crystal before and after shattering during the phase transition.
	}
\end{figure}

\subsection{Physical properties}
\label{sec:Results_PhysicalProperties}

%\textbf{Magnetization} 
\label{sec:Results_chiT}
Figure\,\ref{fig:PhysicalProperties} presents the low temperature physical properties of BaFe$_2$Al$_9$ and BaCo$_2$Al$_9$. In Fig.\,\ref{fig:PhysicalProperties}a, the Co compound shows a relatively small, paramagnetic susceptibility, $\chi$, with weak temperature dependence across the whole temperature range. This is consistent with the Pauli paramagnetic response of a metal.\cite{Blundell2001_MagnetismInCondensedMatter,Kittel2004_SolidStatePhysics} 

BaFe$_2$Al$_9$ has a similar flat magnetic response at high temperatures but the susceptibility rapidly decreases around 100\,K on cooling. Below this, the $\chi$ is nearly temperature independent once again. This change in magnetic response occurs over a 10\,K range and displays a 10\,K hysteresis between heating and cooling. Both features are suggestive of a first-order phase transition in BaFe$_2$Al$_9$ at around 100\,K. Neither compound has a significant Curie-Weiss contribution suggestive of local magnetic moments. The weak peak in both curves near 20\,K is not present in all samples and is likely an impurity.

%\textbf{Resistance} 
\label{sec:Results_RT}
In Fig.\,\ref{fig:PhysicalProperties}b the electrical resistance of both compounds exhibit metallic behavior with resistance generally increasing with temperature. The data for BaFe$_2$Al$_9$ displays a conspicuous 60\% jump around 100\,K on cooling between two metallic regimes. This hysteretic feature coincides with the first order transition observed in the magnetization measurement. Although the orientation of the small BaFe$_2$Al$_9$ crystal used for this measurement is unknown, another sample had a resistivity of 144\,\textmu$\Omega$\,cm along the $c$-axis at 300\,K.

BaCo$_2$Al$_9$ shows uninterrupted metallic resistance down to base temperature with resistivity along the $c$-axis of 2.3 and 16\,\textmu$\Omega$\,cm at 2 and 300\,K, respectively. This results compares well with the resistance behavior reported by \citeauthor{Ryzynska2020_SingleCrystalPropertiesMCo2Al9} although we observe a smaller residual resistance ratio of 6.7 instead of 10.\cite{Ryzynska2020_SingleCrystalPropertiesMCo2Al9} Neither BaFe$_2$Al$_9$ or BaCo$_2$Al$_9$ display superconductivity above 1.85\,K.

%\textbf{Heat Capacity} 
\label{sec:Results_CpT}
Figure \ref{fig:PhysicalProperties}c presents the low-temperature specific heat capacity of the two compounds. BaCo$_2$Al$_9$ exhibits a relatively featureless heat capacity curve. The BaFe$_2$Al$_9$ data have a slightly larger $C_p$ across the whole range and a subtle change in slope near 100K. A first order transition would be expected to produce a spike at this temperature corresponding to the latent heat. Instead, we find a few data points with elevated values of $C_p$. These are likely due to the portions of the sample transforming at different times. The 10\,K transition width observed in magnetic susceptibility also suggests an inhomogeneous transformation. This is likely why a single, sharp peak expected for a first-order phase transition is not observed near 100\,K in the BaFe$_2$Al$_9$ specific heat data.

The inset of Fig.~\ref{fig:PhysicalProperties}c shows fits to the lowest temperature specific heat data to estimate the electronic heat capacity term, $\gamma$. We estimate values of 19.4(1) and 4.44(2)\,mJ/molF.U.\,K$^2$ for the BaFe$_2$Al$_9$ and BaCo$_2$Al$_9$, respectively. \citeauthor{Ryzynska2020_SingleCrystalPropertiesMCo2Al9} calculated a $\gamma$ of BaCo$_2$Al$_9$ to be 7.94 mJ/molF.U.\,K$^2$. Our estimated $\gamma$ of the Fe compound is 4.4 times that of the Co variant suggesting a larger electronic density of states in the low temperature phase\cite{Kittel2004_SolidStatePhysics}. In contrast, the low temperature magnetic susceptibilities of the two compounds (Fig.~\ref{fig:PhysicalProperties}a) suggest that they have comparable density of states below 100\,K. Pauli paramagnetic and Landau diamagnetic contributions are proportional to the density of states and should be dominant in these two metals.\cite{Kittel2004_SolidStatePhysics} Therefore, similar $\chi$ values imply similar density of states. The origin of this contradiction is not clear, but the enhanced value of $\gamma$ in BaFe$_2$Al$_9$ might arise from electronic correlations.

%\textbf{Lattice vs temperature} 
\label{sec:Results_acVsT}
Figure \ref{fig:PhysicalProperties}d presents the temperature dependence of the lattice parameters of BaFe$_2$Al$_9$ and BaCo$_2$Al$_9$ determined by powder x-ray (small symbols) and neutron diffraction (larger symbols). BaCo$_2$Al$_9$ shows a relatively isotropic contraction of the lattice by 0.2\% for both $a$ and $c$ between 300 and 15\,K. This is in contrast to the anisotropic thermal expansion observed in isostructural BaIr$_2$In$_9$.\cite{Calta2015_BaIr2In9ThermalExpansion} The thermal expansion of BaFe$_2$Al$_9$ shows a similar uniform lattice contraction down to 100\,K. The lattice parameters change discontinuously at the transition observed in magnetization and resistance measurements. $a$ abruptly increases by 0.5\% and $c$ shrinks by 1.5\% for a net volume contraction of 0.5\%. Coexistence of the high and low temperature phases is observed between 80 and 110\,K in both the x-ray and neutron experiments. Phase coexistance and the jump in the lattice parameters provide further evidence that the transition in BaFe$_2$Al$_9$ is truly first order.

%\textbf{Comment on isotropic thermal expansion and compare to BaIr2In9.} \cite{Calta2015_BaIr2In9ThermalExpansion}\linebreak
%a 0.71(2)x10$^{-5}$ 1/K
%c 2.33(2)x10$^{-5}$ 1/K\linebreak
%BaFe$_2$Al$_9$ thermal expansion (cryoXRD WM0103 fit 140-300K normalized to 300K)\linebreak
%a 1.20(3)x10$^{-5}$ 1/K
%c 1.36(6)x10$^{-5}$ 1/K\linebreak
%BaCo$_2$Al$_9$ thermal expansion (cryoXRD WM0074 fit 100-300K normalized to 300K)\linebreak
%a 0.91(5)x10$^{-5}$ 1/K
%c 1.07(6)x10$^{-5}$ 1/K

%\textbf{Crystals shatter} 
\label{sec:Results_XtalShatter}
This inhomogeneous first-order transformation has catastrophic consequences for crystals of BaFe$_2$Al$_9$. Samples cooled through 100\,K physically shatter. Crystals larger than 1\,mm fragment into 0.01 - 0.3 mm pieces as depicted by the insets of Fig.~\ref{fig:PhysicalProperties}d. This can be accomplished by dropping them in liquid nitrogen. This self-destructive behavior likely stems from the large, 1.5\% strain at the transition. Rigid inclusions and an inhomogeneous transformation will develop significant stresses in samples causing them to fail. This unfortunate feature of BaFe$_2$Al$_9$ complicated physical property measurements. Samples either had to be held together mechanically or had to be small enough that they survived the transformation.

\begin{figure}
	\includegraphics[width=7in]{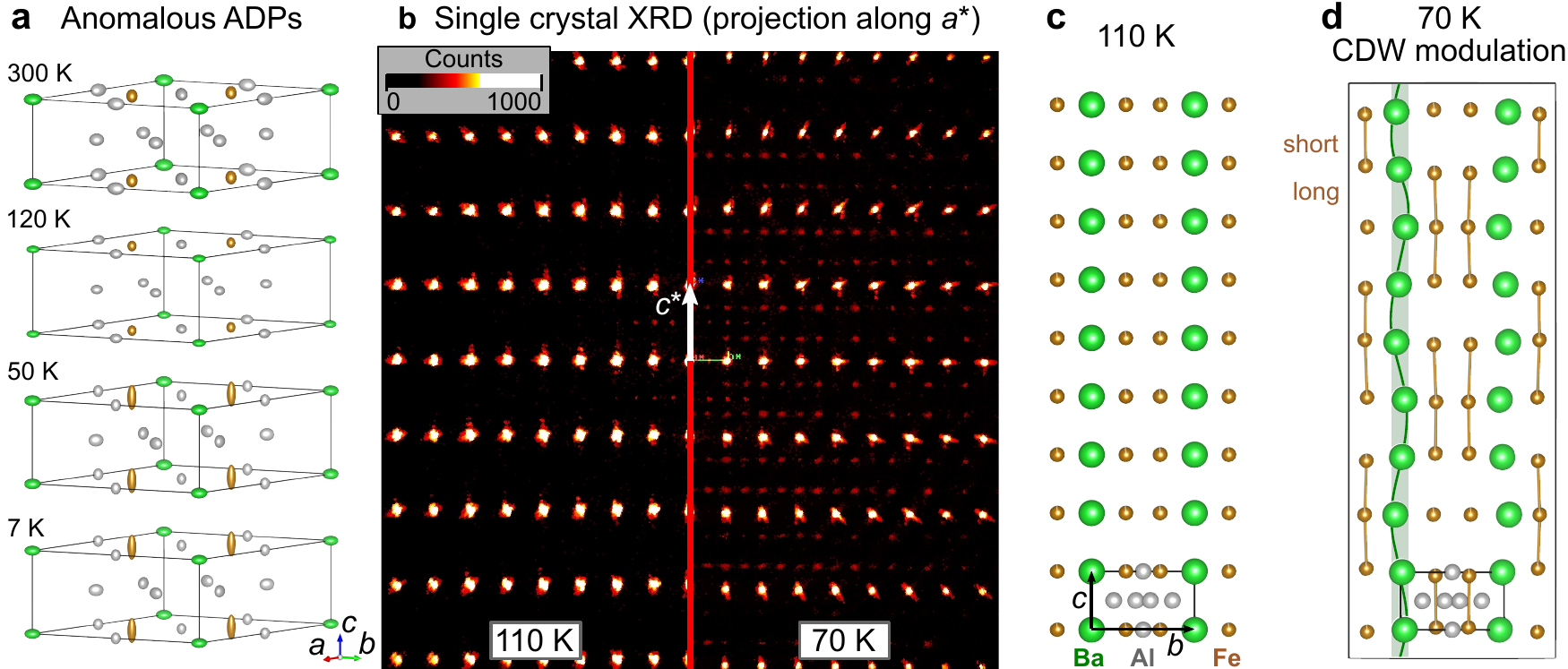}
	\caption{\label{fig:CDW_data} 
		Evidence for a charge density wave (CDW) in BaFe$_2$Al$_9$. 
		(\textbf{a}) Atomic displacement parameters (ADP) from powder neutron diffraction (represented by 99\% displacement ellipsoids) are anomalously large below 100\,K suggesting a change in crystal structure.
		(\textbf{b}) Single crystal x-ray diffraction data with all intensity projected along $a^*$ showing new super-lattice peaks at 70\,K. The origin is at the center. Peaks at half-integer $L$ near the origin at both temperatures are half-wavelength reflections. 
		(\textbf{c}) Projection of BaFe$_2$Al$_9$ structure along the hexagonal $a$ direction. (\textbf{d}) Depiction of the CDW structure showing Ba and Fe position modulations. Vertical links between Fe atoms denote shorter than average Fe-Fe distances to emphasize their vertical displacements.
	}
\end{figure}

\subsection{Evidence for charge density wave}
\label{sec:Results_CDW}

The temperature-dependent physical properties presented in Fig.~\ref{fig:PhysicalProperties} suggest that a first-order phase transition occurs in BaFe$_2$Al$_9$ around 100\,K. Powder x-ray and neutron diffraction measurements did not reveal any evidence for new diffraction peaks, measurable deviations from hexagonal symmetry or magnetic order. Refinement of the neutron diffraction data revealed a dramatic increase in Fe and Ba atomic displacement parameters (ADPs) on cooling through 100\,K (Fig.~\ref{fig:CDW_data}a and supplemental information). This increase was notably anisotropic; Fe atoms displace more along the [001] direction and Ba in the $ab$-plane. This observation signals a potential lattice modification in the low temperature phase.

%\textbf{Single crystal XRD} 
\label{sec:Results_scXRD}
We turn to single crystal x-ray diffraction to expose the nature of this phase transition. Figure \ref{fig:CDW_data}b presents all the measured intensity in reciprocal space projected along the $\bm{a}^*$ direction (i.e. projected onto the ($-H\ 2H\ L$) reciprocal plane). At 110\,K (left side) a rectangular array of peaks is observed consistent with the room-temperature $P6/mmm$ unit cell. At 70\,K, new peaks appear signaling a change in lattice periodicity. These superlattice peaks are on the order of 1/300th the intensity of the primary reflections and can be indexed by three symmetry-related wave vectors: \textonehalf\,0\,$k_z$, -\textonehalf\,\textonehalf\,$k_z$, and 0\,-\textonehalf\,$k_z$. We determined the incommensurate index, $k_z =$ 0.3020(9)\,r.l.u.~at 70\,K. These peaks signal the presence of a charge density wave (CDW) modulation in the low temperature phase of BaFe$_2$Al$_9$.

Using ISODISTORT, we determined that $P6/mmm$ has 4 irreducible representations with wave-vector \textonehalf\,0\,$k_z$. \cite{ISODISTORT_6.7.2,Campbell2006_ISODISPLACE} Among these, the $U_4$ displacement modes were consistent with anomalous Fe and Ba ADPs noted above and gave the best agreement with the superlattice intensities using Jana2020.

%\textbf{Modulated structure} 
\label{sec:Results_ModStructCartoon}
Figure \ref{fig:CDW_data}c shows the refined 110\,K hexagonal structure viewed along the $a$ direction and Fig.~\ref{fig:CDW_data}d shows the corresponding view of our best model of the modulated CDW structure at 70K. This is a commensurate approximate version of the CDW with 20-times larger, C-centered orthorhombic unit cell (thin black rectangle) corresponding to the super-lattice wave-vector. First, note that straight columns of barium atoms at 110\,K along the [001] direction develop a sine-modulated, horizontal displacement with a period of $\frac{10}{3} c$. The Fe atoms also have modulated displacements primarily along [001] leading to varying Fe-Fe distances. Lines connect Fe atoms that are closer than average emphasizing this change. These displacements correspond to the directions with large Fe and Ba ADPs in Fig.~\ref{fig:CDW_data}a at low temperature.

We are confident that the dominant distortion mode is similar to the one depicted in Fig.~\ref{fig:CDW_data}d and transforms at the $U_4$ irreducible representation.
%We are confident that the dominant distortion mode transforms as the $U_4$ irreducible representation and is similar to the one depicted in Fig.~\ref{fig:CDW_data}d.
This model best explains the ADPs of the refined average  structures and intensity of the superlattice peaks. Unfortunately, our x-ray data does not provide a complete refinement of the modulated structure. Attempts to refine the modulated structure in Jana2020 yielded confident modulation parameters for the heavier Ba and Fe atoms but not for the two Al sites. Including these parameters did not give consistent or stable solutions and negative ADP parameters signaled that this model is insufficient to completely describe the data.

\begin{figure}
	\includegraphics[width=3.33in]{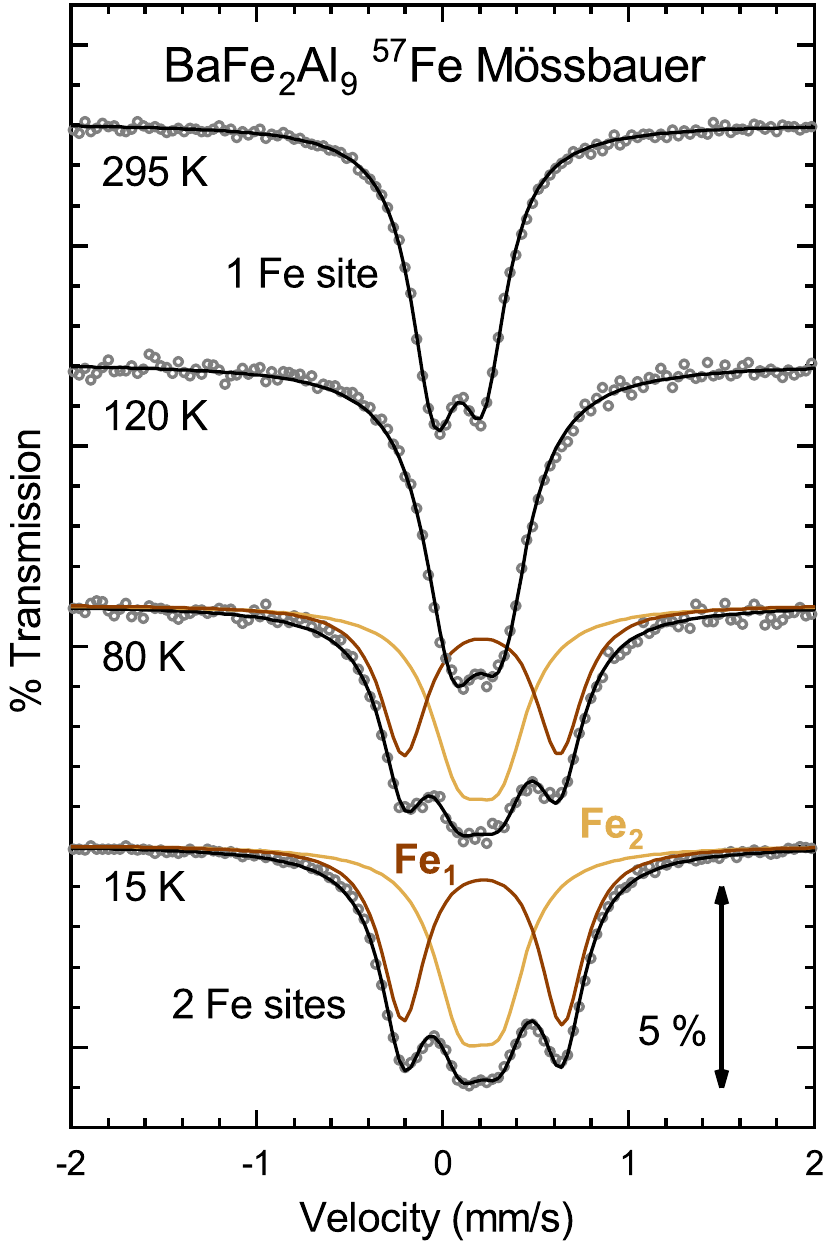}
	\caption{\label{fig:Mossbauer} 
		$^{57}$Fe M\"ossbauer spectroscopy of powdered BaFe$_2$Al$_9$ shows an absorption doublet from the single Fe site in hexagonal BaFe$_2$Al$_9$ at 120 and 295\,K. In the low temperature CDW phase (80 and 15\,K) the spectra are best fit with two doublets corresponding to two distinct Fe sites.
	}
\end{figure}

%\textbf{M\"ossbauer} 
\label{sec:Results_Moss}
In addition to our partial structure refinement, $^{57}$Fe M\"ossbauer spectroscopy indicates that our CDW description in Fig.~\ref{fig:CDW_data}d is missing something. Figure \ref{fig:Mossbauer} presents the M\"ossbauer spectra for BaFe$_2$Al$_9$ above and below the first order transition. At 295 and 120\,K, a single absorption doublet is observed generated by the single iron site in the room-temperature structure. The small asymmetry suggests a minor preferential orientation. The spectra change abruptly on cooling through the transition and four peaks can be resolved. 
This spectrum is best modeled by a pair of doublets as show. These correspond to two iron sites with different quadrupole splitting values and isomer shifts with equal population.
%This spectrum is best modeled by a pair of doublets from two, equal populations of Fe sites with different quadrupole splitting values and isomer shifts as shown. 
In short, the low temperature phase of BaFe$_2$Al$_9$ has two types of iron atoms with different local environments. None of the measured spectra showed evidence of magnetic hyperfine splitting. Based on magnetic susceptibility, neutron diffraction, and this M\"ossbauer result, BaFe$_2$Al$_9$ does not appear to undergo magnetic ordering.

First, this M\"ossbauer evidence for two Fe sites at low temperature supports our identification of a structural change below 100\,K. 
However, our diffraction result is not immediately consistent with two Fe sites. The first-order super-lattice peaks observed only provide information on the first harmonic of the position modulations. The simplest (sinusoidal) modulation this describes would give a distribution of iron site environments with a range of M\"ossbauer parameters, not two distinct Fe-site populations. The M\"ossbauer spectra are not well fit by such a distribution of quadrupole splittings and isomer shifts (see supplemental material). 

There are likely higher-order harmonics to the CDW modulation which generate the two distinct Fe atom environments suggested by M\"ossbauer spectroscopy. These additional modulation components should generate higher-order super-lattice peaks\cite{Boucher1996_IncommensurateStructureTaGeTe2}. We suspect these were too weak for us to resolve with the single crystal diffractometer. Between an inconclusive x-ray refinement and the M\"ossbauer result, the CDW structure of BaFe$_2$Al$_9$ presented in Fig.~\ref{fig:CDW_data}d is not yet a complete description of the low temperature structure.

\section{Discussion}

The temperature dependent physical properties in Fig.~\ref{fig:PhysicalProperties} reveal a first-order phase transition in BaFe$_2$Al$_9$ around 100\,K which tends to shatter the crystals. Superlattice peaks observed with single crystal XRD reveal that the low temperature phase is a charge density wave. No such transition is observed in BaCo$_2$Al$_9$. Electronic structure calculations provide one clue to why a CDW appears in the Fe and not the Co compound; the filling of the transition-metal d-orbitals.

\begin{figure}
	\includegraphics[width=17.78cm]{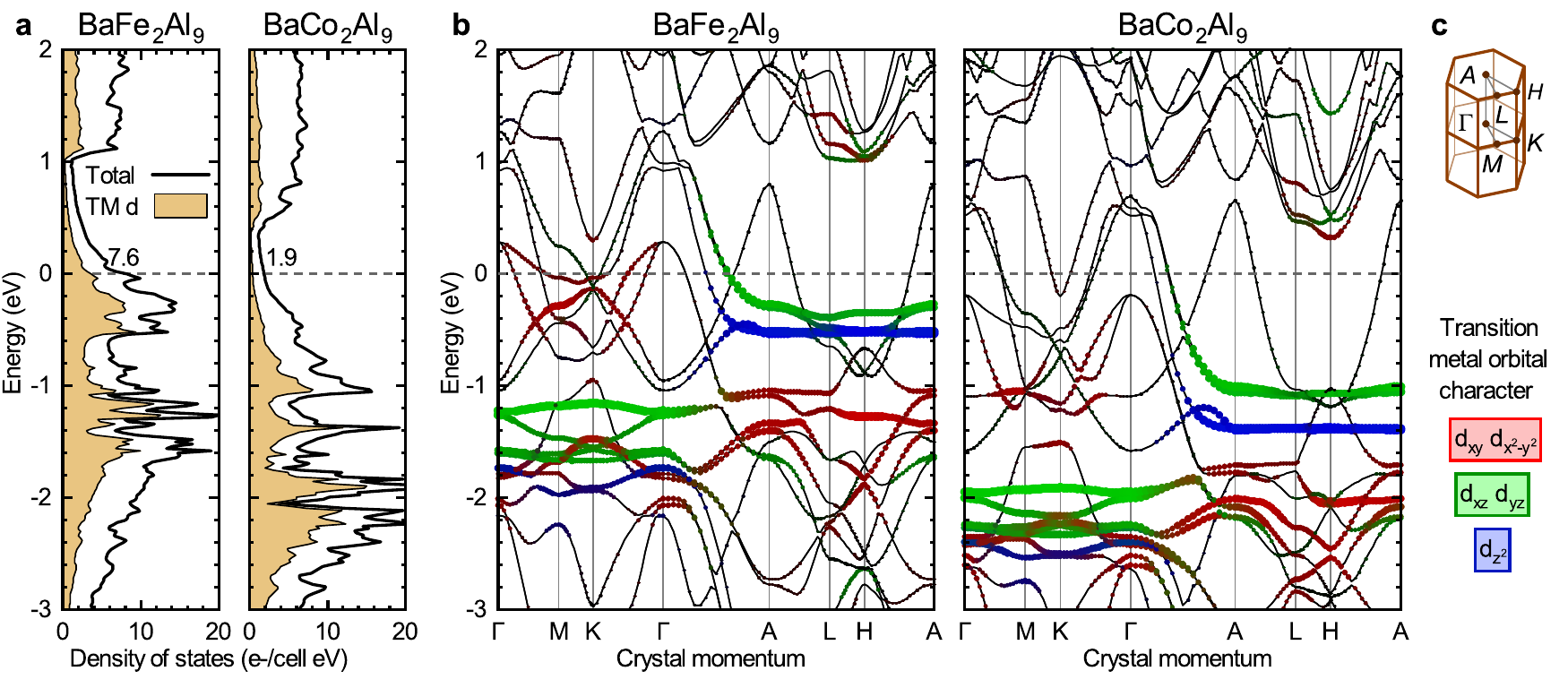}
	\caption{\label{fig:DFT} 
		Electronic structures of BaFe$_2$Al$_9$ and BaCo$_2$Al$_9$ from density functional theory. (\textbf{a}) The electronic density of states for both compounds. The yellow fill represents the transition metal (TM) d-orbital contribution. The calculated density of states in electrons/unit cell eV sit above the Fermi level (dashed line). (\textbf{b}) The electronic band structure of the two aluminides. Thicker lines have more d-orbital character and the color represents contributions of specific d-orbitals. Note that the thicker d-bands are only partially filled in BaFe$_2$Al$_9$. (\textbf{c}) Brillouin zone of BaFe$_2$Al$_9$ with momentum point labels.
	}
\end{figure}

\subsection{Electronic structure calculations}
\label{sec:Discussion_ElectronicStructure}

Figure \ref{fig:DFT} presents the calculated electronic structure of BaFe$_2$Al$_9$ and BaCo$_2$Al$_9$ from DFT. Panel \ref{fig:DFT}a provides a comparison of the density of states between the two compounds. The shaded region represents the d-orbital contribution at each energy. The d-states are shifted to lower energies in the Co compound dramatically reducing the density of states at the Fermi energy (dashed line). 
	
The band structure in Fig.~\ref{fig:DFT}b shows more detail of these changes.
The thicker bands have more transition-metal d-orbital character and the color denotes the contribution from each d-orbital variety. Thinner bands are generally dominated by Al s- and p-orbitals in the depicted energy range.

Note the common features between the two compounds. The thicker d-orbital bands show small dispersion in $\Gamma-M-K$ and $A-L-H$ planes but they are shifted down by 0.7-0.9\,eV in the Co compound. Critically, in BaFe$_2$Al$_9$ several of the bands at the Fermi level (dashed line) have d-orbital character; the red ($d_{xy}$/$d_{x^2-y^2}$) bands in the $\Gamma-M-K$ plane and green ($d_{xz}$/$d_{yz}$) bands on the $\Gamma-A$ line. In the BaCo$_2$Al$_9$, only thin lines representing Al-dominated bands lie at the Fermi level. 

We propose that the partly filled d-bands play an important role in the formation of the CDW in BaFe$_2$Al$_9$. First, this is the most conspicuous difference between the electronic structures in Fig.~\ref{fig:DFT}. Second, our physical property measurements suggest that the CDW only opens a gap in some of the bands. In a canonical, single-band CDW system the structural modulation is heralded by a metal-insulator transition. High-temperature metallic properties give way to semiconducting behavior as a band gap opens on cooling.\cite{Gruner1994_DensityWavesInSolids} In contrast, BaFe$_2$Al$_9$ displays metallic resistance behavior both above and below the transition (Fig.~\ref{fig:PhysicalProperties}b). 

In BaFe$_2$Al$_9$ many bands cross the Fermi energy. CDW formation likely gaps out some of these bands (maybe the d-bands) which reduces the density of states. This reduction of metallic carriers causes the abrupt increase in the resistance on cooling but the un-gapped bands facilitate metallic resistance vs.~temperature behavior at low temperatures. The magnetic susceptibility data presented in Fig.~\ref{fig:PhysicalProperties}a supports this model. The Pauli paramagnetic contribution to the susceptibility is proportional to the density of states at the Fermi level.\cite{Blundell2001_MagnetismInCondensedMatter} Gap formation on some of the bands by the CDW leads to a smaller density of states and, therefore a weaker Pauli paramagnetic response at low temperature. 

We believe the role of the d-orbitals in the CDW we propose is testable. Once the modulated structure of the CDW in BaFe$_2$Al$_9$ is determined, the electronic structure can be calculated for the superstructure. Based on our model, we would anticipate that Fe-dominated states would be gapped at the Fermi level and bands from the Al-network would not. This observation would emphasize the relative involvement of these d-orbital states in the structural modulation.

\citeauthor{Vajenine1998_MagicElectronCountsAeM2Al9} noted something curious about this BaFe$_2$Al$_9$ family of compounds; the transition metal is the most electronegative constituent\cite{Vajenine1998_MagicElectronCountsAeM2Al9}. They suggest that Fe and Co will have filled d-orbitals and the remaining electrons are distributed on the Al-network. 
Our DFT results reveal that this is not quite the case in BaFe$_2$Al$_9$. As we noted above, the d-bands have unfilled states. Their argument still has merit in that the additional electrons per formula unit (F.U.) in BaCo$_2$Al$_9$ preferentially fill the d-bands. The thinner bands, with strong Al character, shift less than the d-bands between the two compounds in Fig.~\ref{fig:DFT}b. Al-bands fall by 0.15-0.4\,eV while the d-bands fall by 0.7-0.9\,eV. The smaller shift of Al-network bands is exemplified by the thin hole band above the Fermi energy at the $A$ point in both compounds. 

This disparate energy shift of the d- and Al-bands suggests that filled d-orbitals are indeed favored in this family of compounds. Unfilled d-orbitals in BaFe$_2$Al$_9$ may explain its abrupt volume change at the CDW transition. To fill the Fe d-orbitals, the electrons would have to come from the Al-network. Reducing the filling of the Al-bands would modify network bonding leading to the changes of the lattice parameters presented in Fig.~\ref{fig:PhysicalProperties}d.

%The abrupt volume change of BaFe$_2$Al$_9$ at the CDW transition might be caused by the readjustment of the Al-network after relinquishing electrons to fill the Fe d-orbitals.

%The disparate energy shifts of the d-bands and Al-bands confirm from the unequal distribution of the additional electrons in the Co compound suggested by \citeauthor{Vajenine1998_MagicElectronCountsAeM2Al9}.

%Co  suggested by \citeauthor{Vajenine1998_MagicElectronCountsAeM2Al9}. Maybe filled d-bands are favored at low temperature in BaFe$_2$Al$_9$ resulting in the smaller volume CDW modulated structure.

The archetypal CDW compounds are characterized by weakly coupled chains or sheets such as NbSe$_3$, K$_{0.3}$MoO$_3$, the $R$NiC$_2$-family, K$_2$Mo$_{15}$Se$_{13}$, NbSe$_2$, and (K,Ba)AgTe$_2$.\cite{Gruner1994_DensityWavesInSolids,Steiner2018_SingleCrystalStudyCDW-LuNiC2,Roman2018_PhaseDiagramOfRNiC2-Family,Candolfi2020_CDW+SC_K2Mo15Se19,Gourdon2000_KBaAgTe2-CDWstructure} CDWs in intermetallic compounds with 3D structural networks, like BaFe$_2$Al$_9$, are notably less common. Examples include a few transition metal silicides, germanides, stannides, and chalcogenides such as $R_2M_3$Si$_5$,\cite{Ramakrishnan2020_CDWinEr2Ir3Si5,Singh2005_CDWinLu2Ir3Si5,Sangeetha2012_CDWdopedLu2Ir3Si5} $R_2M_3$Ge$_5$,\cite{Bugaris2017_CDW-R2Ru3Ge5} and $R_5M_4$Si$_{10}$\cite{Kuo2001_ThermalPropCDW-Lu5Ir4Si10,Kuo2003_IonicSizeCDW-R5Ir4Si10,Lue2002_HysteresisOfCDW-Lu5Rh4Si10,Yang1991_PhaseTransistions-Lu5Rh4Si10}, the Sr$_3$Ir$_4$Sn$_{13}$-family\cite{Klintberg2012_QuantumPhaseTransititionSr3Ir4Sn13,Welsch2019_CDW-La3Co4Sn13} and Ir$_2$In$_8$Se \cite{Khoury2020_Ir2In8Q-DiracSemimetal+StructuralModulation}. Few compounds show first-order CDW transitions like that in BaFe$_2$Al$_9$. Examples include Lu$_2$Ir$_3$Si$_5$, Er$_2$Ir$_3$Si$_5$ \cite{Singh2005_CDWinLu2Ir3Si5,Ramakrishnan2020_CDWinEr2Ir3Si5} and light rare-earth variants of $R$Pt$_2$Si$_2$ \cite{Nagano2013_CDW+SC-RPt2Si2}. The first two compounds show similar jumps in resistance vs temperature with low temperature metallic behavior like BaFe$_2$Al$_9$. Although charge order has been identified in two iron oxides (LuFe$_2$O$_4$ \cite{Ikeda2005_FerroelectricityFromFeValanceOrderInLuFe2O4} and magnetite \cite{Walz2002_TheVerweyTransition}), BaFe$_2$Al$_9$ is the first example of a CDW in an Fe-based compound we are aware of without magnetism.

%1D chains of atoms hold an important role in understanding the origin of CDW modulations.\cite{Gruner1994_DensityWavesInSolids} The $c$-axis Ba chains in BaFe$_2$Al$_9$ likely play a critical role in CDW formation despite DFT indicating that they do not contribute to the bands at the Fermi energy. These chains sit in channels formed by the aluminum network (see Fig.~\ref{fig:Structure}) and transverse displacements of these heavy atoms (like those in Fig.~\ref{fig:CDW_data}d) should constitute low energy phonon modes. If a favorable electronic instability is present, these displacement modes provide the lattice modulation piece of a CDW instability. This suggests that intermetallic network compounds with channels hosting linear chains of large, heavy atoms are good candidates to search for low temperature modulated structures. First-principle calculations of phonons might provide additional guidance.

%\textbf{Open questions} 
\label{sec:Discussion_OpenQs}
Identifying a CDW ground-state in BaFe$_2$Al$_9$ begs two key questions. First, iron is renowned for its magnetism and partially filled d-orbitals are critical ingredients for magnetic order.\cite{Blundell2001_MagnetismInCondensedMatter} This is particularly true for 3d transition-metal compounds. Therefore, it is somewhat surprising to find a CDW in BaFe$_2$Al$_9$ despite its nearly filled Fe d-bands. Why is a charge modulation favored over magnetism in this compound?

Second, the majority of charge density wave transitions are second-order or weakly first order.\cite{Gruner1994_DensityWavesInSolids} In contrast, the CDW in BaFe$_2$Al$_9$ appears at a strongly first-order transition with a 1.5\% shrinkage along $c$. Why is the CDW transition first order in this compound? What is the role of the large lattice strain that accompanies it?

In general, the mechanism that drives CDW transitions is not entirely clear.\cite{Gruner1994_DensityWavesInSolids,Zhu2017_MisconceptionsChargeDensityWaves} This is especially true in materials with relatively 3D electronic structures and many Fermi surfaces, like BaFe$_2$Al$_9$. Although Fermi surface nesting is frequently discussed as driving CDW formation, this argument is strongest in materials with particularly 1D character. Electron correlations and wavelength-dependent electron-phonon coupling may be more important in 3D materials like BaFe$_2$Al$_9$.

This compound offers a unique opportunity for chemical tuning to explore the origin of its CDW order. First, unlike the more common second-order CDW transition, the first-order transition produces a strong signal for tracking changes. Magnetic susceptibility, resistance, and particularly the lattice parameters all show dramatic discontinuities. This will allow us to easily track the phase transition as we modify the material. Next, BaFe$_2$Al$_9$ belongs to a family with some compositional flexibility.\cite{Vajenine1998_MagicElectronCountsAeM2Al9} As a result, this compound offers several opportunities for chemical substitution. Isovalent doping of Sr for Ba, Ru for Fe, or Ga for Al allow tuning of the lattice size and disorder. The role of band filling could be explored by replacing Fe with other transition metals. Overall, BaFe$_2$Al$_9$ has a wonderful potential for tuning its conspicuous first-order CDW transition to learn what drives it. 

\label{sec:Discussion_Summary}

%\textbf{Summary} 
In summary, we grew crystals of BaFe$_2$Al$_9$ and BaCo$_2$Al$_9$ from aluminum flux and characterized their temperature-dependent physical properties. These results reveal no transitions in BaCo$_2$Al$_9$ and a first-order phase transition in the Fe compound near 100\,K. Diffraction results reveal a discontinuous change in lattice parameters across the transition. This strain is large enough to physically shatter the crystals. Single crystal x-ray diffraction revealed the low-temperature phase in BaFe$_2$Al$_9$ to be a charge density wave (CDW) with modulated Fe and Ba positions. Comparing the electronic structures of BaFe$_2$Al$_9$ and BaCo$_2$Al$_9$ hints that the CDW in the former might be associated with Fe d-bands at the Fermi energy. BaFe$_2$Al$_9$ offers many opportunities for chemical tuning to answer some outstanding questions about its unusual CDW ground state.

\begin{acknowledgement}
	\label{sec:Acknowledgment}
	
	We would like to thank Anna B\"ohmer, Andreas Kreyssig, Joe Paddison, Andrew May, Jiaqiang Yan, Gordon Miller, Tom Roberson and Rob Moore for their discussions and insights. We would also like to thank V\'aclav Pet\v{r}\'i\v{c}ek for his assistance with the modulated structure which required an early copy of Jana2020.
	
	Research supported by the U. S. Department of Energy, Office of Science, Basic Energy Sciences, Materials Sciences and Engineering Division (under contract number DE-AC05-00OR22725). GDS was supported as part of the Energy Dissipation to Defect Evolution (EDDE), an Energy Frontier Research Center funded by the US Department of Energy, Office of Science, Basic Energy Sciences under Contract Number DE-AC05-00OR22725.This research used resources at the Spallation Neutron Source, a DOE Office of Science User Facility operated by Oak Ridge National Laboratory.

\end{acknowledgement}

\begin{suppinfo}
\label{sec:suppinfo}

%A listing of the contents of each file supplied as Supporting Information should be included. For instructions on what should be included in the Supporting Information as well as how to prepare this material for publications, refer to the journal's Instructions for Authors.

The following files are available free of charge.
\begin{itemize}
	\item CDW\_in\_BaFe2Al9\_SI\_02.pdf : Details of neutron powder diffraction refinements and M\"ossbauer fits
	\item BaFe2Al9\_neutron\_300K\_submit.cif : Refined structure of BaFe$_2$Al$_9$ at 300\,K using neutron diffraction
\end{itemize}

\textbf{Supplemental Information}

\textbf{Neutron powder diffraction refinement}
\label{sec:suppinfo_NeutronRefine}

\begin{figure}
	\includegraphics[width=7in]{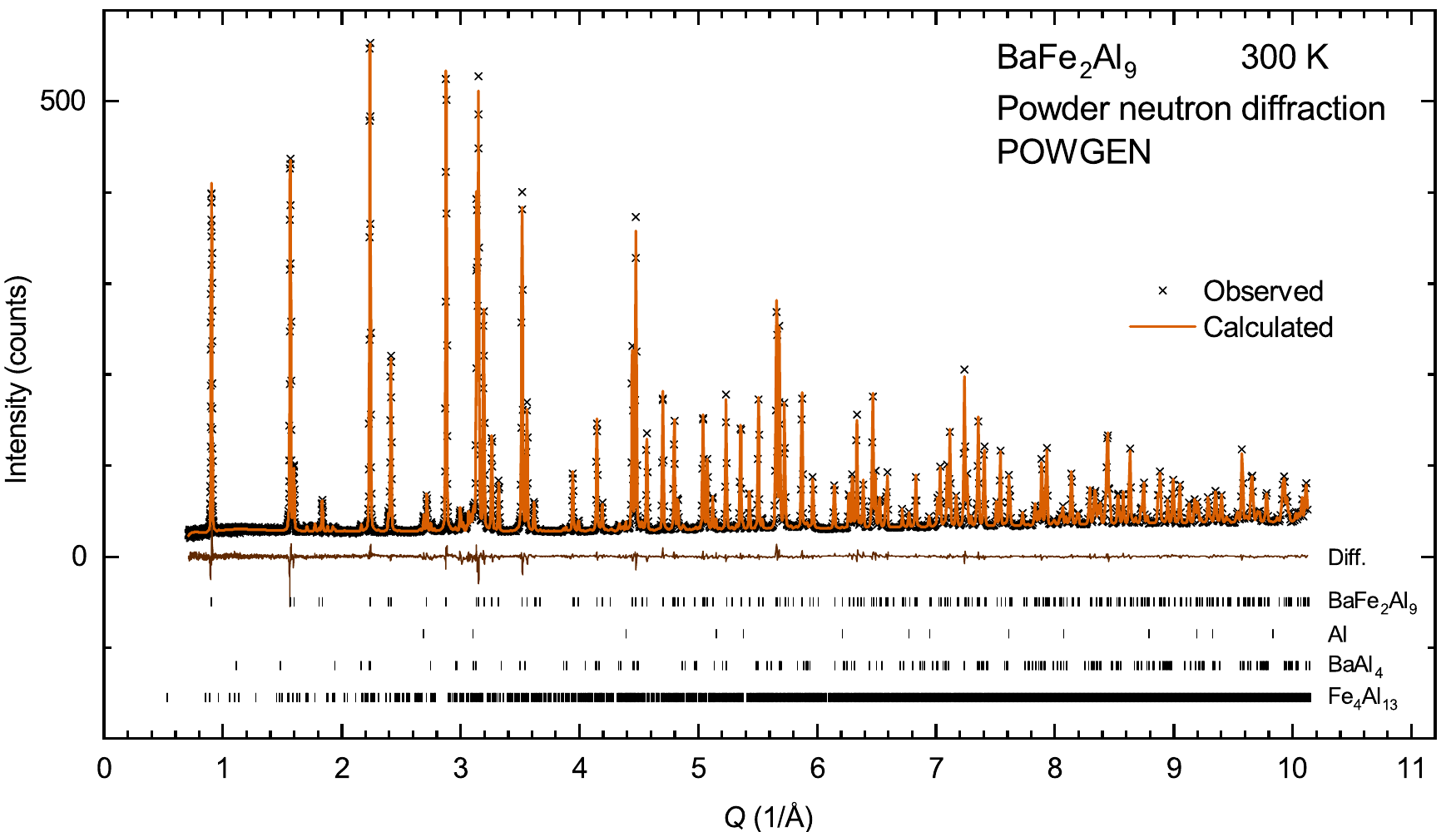}
	\caption{\label{fig:SI_NeutronFit} 
		Fit of the 300 K neutron powder diffraction data taken at POWGEN.
	}
\end{figure}

Figure \ref{fig:SI_NeutronFit} depicts the fit to the powder neutron diffraction of \BaFeAl\ taken at 300\,K. We were not selective with the crystals chosen to grind into the 1.83\,g sample. As a result, this measurement represents a worst-case scenario for impurities. The refinement indicates 92.2(3)\,wt\% \BaFeAl\, 1.14(7)\,wt\% Al metal, 2.62(19)\,wt\% BaAl$_4$, and 4.00(12)\,wt\% Fe$_4$Al$_{13}$.

Tables \ref{tbl:NeutronRefDetails} presents the details of the structure refinement from the powder neutron diffraction data of \BaFeAl\ above and below the charge density wave (CDW) transition. Although the compound has a CDW modulation at 50 and 7\,K the same structure model was used for all temperatures. This was done because no super-lattice peaks are observed in the neutron data to refine a modulated structure. We were able to infer the presence of a CDW modulation in the average structure based on the atomic parameters in Tables \ref{tbl:NeutronRefAtoms} and \ref{tbl:NeutronRefADPs}. As discussed in the main text, the anisotropic atomic displacement parameters (ADPs) show an anomalous temperature dependence below 100\,K consistent with the CDW model we propose.

Our refinement indicated that the Fe site in \BaFeAl\ mostly occupied with only a few percent vacancies. This reduced site occupancy was also observed when x-ray data were refined.

\begin{table}
	\caption{Neutron powder refinement details from POWGEN time of flight data.}
	\label{tbl:NeutronRefDetails}
	\begin{tabular}{l| c c c c}
		Chemical formula &\multicolumn{4}{c}{BaFe$_{1.92}$Al$_9$}\\
		Formula wt. (g/molF.U.) &\multicolumn{4}{c}{487.383}\\
		Crystal system &\multicolumn{4}{c}{hexagonal}\\
		Space group &\multicolumn{4}{c}{$P6/mmm$ (191)}\\
		Temperature (K) &300 &120 &50 &7\\
		$a$ (\AA) &8.018667(16) &8.006172(15) &8.04357(3) &8.04363(3)\\
		$c$ (\AA) &3.934816(14) &3.929166(13) &3.86933(3) &3.86835(3)\\
		$V$ (\AA$^3$) &219.1086(10) &218.1127(9) &216.8024(19) &216.7509(18)\\
		$Z$ &\multicolumn{4}{c}{1}\\
		Density calculated (g/cm$^3$) &3.7277 &3.7447 &3.7674 &3.7683\\
		wavelength range (\AA) &\multicolumn{4}{c}{0.97 - 2.033}\\
		F(000) &494.718 &494.718 &482.026 &494.718\\
		Time of flight range (ms) &\multicolumn{4}{c}{11.3 - 315}\\
		No. of variables &\multicolumn{4}{c}{46}\\
		RF(obs) &1.15 &1.17 &2.7 &2.65\\
		RFw(obs) &1.58 &1.69 &4.4 &4.15\\
		Goodness of fit &2.8 &3 &4.58 &4.27\\
		Absorption &0.077(2) &0.077(2) &0.078(4) &0.074(3)\\
		
	\end{tabular}
\end{table}

\begin{table}
	\caption{Neutron refinement \BaFeAl\ atomic parameters}
	\label{tbl:NeutronRefAtoms}
	\begin{tabular}{l| c c c c c c }
		Atom &Site &$x$ &$y$ &$z$ &$U_{aniso.}$ &Occupancy\\
		\hline
		&\multicolumn{6}{c}{300 K}\\
		Al1 &3$f$ &\textonehalf &0 &0 &0.0099(2) &1\\
		Al2 &6$m$ &0.21481(4) &0.42961(8) &\textonehalf &0.0089(2) &1\\
		Ba &1$a$ &0 &0 &0 &0.0126(3) &1\\
		Fe &2$c$ &$\frac{1}{3}$ &$\frac{2}{3}$ &0 &0.0067(1) &0.984(3)\\
		\hline
		&\multicolumn{6}{c}{120 K}\\
		Al1 &3$f$ &\textonehalf &0 &0 &0.0059(2) &1\\
		Al2 &6$m$ &0.21484(4) &0.42968(8) &\textonehalf &0.0055(2) &1\\
		Ba &1$a$ &0 &0 &0 &0.0064(2) &1\\
		Fe &2$c$ &$\frac{1}{3}$ &$\frac{2}{3}$ &0 &0.0045(1) &0.985(3)\\
		\hline
		&\multicolumn{6}{c}{50 K}\\
		Al1 &3$f$ &\textonehalf &0 &0 &0.0055(4) &1\\
		Al2 &6$m$ &0.21355(7) &0.42711(13) &\textonehalf &0.0077(3) &1\\
		Ba &1$a$ &0 &0 &0 &0.0101(4) &1\\
		Fe &2$c$ &$\frac{1}{3}$ &$\frac{2}{3}$ &0 &0.0127(2) &0.988(5)\\
		\hline
		&\multicolumn{6}{c}{7 K}\\
		Al1 &3$f$ &\textonehalf &0 &0 &0.0053(3) &1\\
		Al2 &6$m$ &0.21354(6) &0.42707(12) &\textonehalf &0.0076(3) &1\\
		Ba &1$a$ &0 &0 &0 &0.0093(4) &1\\
		Fe &2$c$ &$\frac{1}{3}$ &$\frac{2}{3}$ &0 &0.0125(2) &0.985(5)\\
		
	\end{tabular}
\end{table}

\begin{table}
	\caption{Neutron refinement \BaFeAl\ atomic displacement parameters}
	\label{tbl:NeutronRefADPs}
	\begin{tabular}{l| c c c c}
		Atom &$U_{11}$ &$U_{22}$ &$U_{33}$ &$U_{12}$\\
		\hline
		&\multicolumn{4}{c}{300 K}\\
		Al1 &0.0128(3) &0.0071(4) &0.0078(4) &0.00356(18)\\
		Al2 &0.0093(2) &0.0105(2) &0.0073(3) &0.00525(12)\\
		Ba &0.0156(3) &0.0156(3) &0.0067(5) &0.00782(17)\\
		Fe &0.00598(15) &0.00598(15) &0.00898(19) &0.00299(7)\\
		\hline
		&\multicolumn{4}{c}{120 K}\\
		Al1 &0.0073(3) &0.0056(3) &0.0043(3) &0.00282(17)\\
		Al2 &0.00589(19) &0.0069(2) &0.0040(3) &0.00344(11)\\
		Ba &0.0083(3) &0.0083(3) &0.0026(4) &0.00414(14)\\
		Fe &0.00389(14) &0.00389(14) &0.00570(17) &0.00195(7)\\
		\hline
		&\multicolumn{4}{c}{50 K}\\
		Al1 &0.0049(4) &0.0045(5) &0.0069(6) &0.0022(3)\\
		Al2 &0.0064(3) &0.0107(4) &0.0073(5) &0.0054(2)\\
		Ba &0.0125(5) &0.0125(5) &0.0052(7) &0.0063(2)\\
		Fe &0.0038(3) &0.0038(3) &0.0305(4) &0.00188(13)\\
		\hline
		&\multicolumn{4}{c}{7 K}\\
		Al1 &0.0048(4) &0.0046(5) &0.0064(5) &0.0023(2)\\
		Al2 &0.0064(3) &0.0108(4) &0.0070(4) &0.00539(19)\\
		Ba &0.0118(4) &0.0118(4) &0.0044(7) &0.0059(2)\\
		Fe &0.0035(2) &0.0035(2) &0.0305(4) &0.00176(12)\\
		
	\end{tabular}
\end{table}

\textbf{M\"ossbauer fits}
\label{sec:suppinfo_Mossbauer}

\begin{figure}
	\includegraphics[width=10cm]{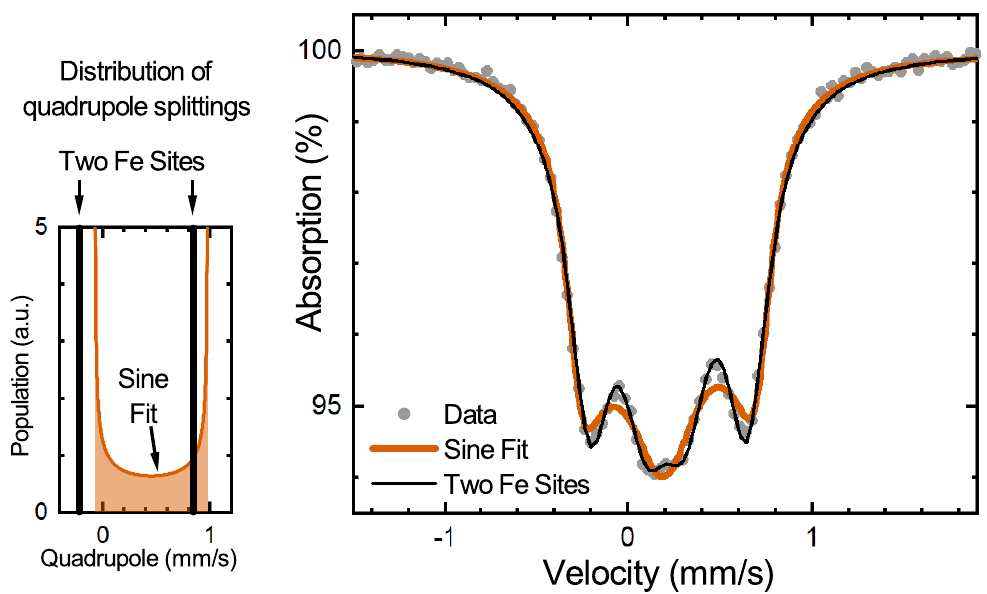}
	\caption{\label{fig:SI_MossbauerFits} 
		Fits of $^{54}$ M\"ossbauer spectra from \BaFeAl\ in the charge density wave phase at 15\,K. The left panel shows the distribution of quadrupole parameters for two Fe sites model (black) and the sine-modulated quadrupole model that best fit the spectra. The right hand plot shows that two-site model (black) fits the measured spectra (gray points) better than the sine modulated model (orange).
	}
\end{figure}

In the results section we note an inconsistency between out M\"ossbauer spectra from \BaFeAl\ and the charge density wave modulation presented in Fig.~3d. Specifically, the observed super-lattice reflections from single crystal x-ray diffraction signal a simple sine-modulation of the atomic positions. This should give a distribution of Fe atom environments with a range of M\"ossbauer parameters. For example, a sine-modulation of the quadrupole parameter would yield the orange distribution of values in the left panel of Fig.~\ref{fig:SI_MossbauerFits}. This contrasts with the discrete pair of values from a fit to the two Fe-site model (thick black line).

We tried to fit the low temperature M\"ossbauer spectra of \BaFeAl\ with such a modulated model. We summed the spectra of 20 sites with sine-modulated M\"ossbauer parameters. The maximum and minimum values of the quadrupole splitting, isomer shift and linewidth were allowed to vary and the best fit is given by the orange curve on the right-hand plot of Fig.~\ref{fig:SI_MossbauerFits}. This fit is distinctly worse than the two iron site model in black presented in the main text.

\end{suppinfo}


%\bibliography{C:/Users/4wm/Documents/References/MeierReferences.bib}% Produces the bibliography via BibTeX.
\providecommand{\latin}[1]{#1}
\makeatletter
\providecommand{\doi}
{\begingroup\let\do\@makeother\dospecials
	\catcode`\{=1 \catcode`\}=2 \doi@aux}
\providecommand{\doi@aux}[1]{\endgroup\texttt{#1}}
\makeatother
\providecommand*\mcitethebibliography{\thebibliography}
\csname @ifundefined\endcsname{endmcitethebibliography}
{\let\endmcitethebibliography\endthebibliography}{}

\end{document}